\documentstyle[b98proc,epsfig]{article}

\def\Journal#1#2#3#4{{#1} {\bf #2}, #3 (#4)}

\def\PLB{{\em Phys.\ Lett.} B}

\def\PRC{{\em Phys.\ Rev.} C}
\def\ZPC{{\em Z. Phys.} C}

\begin{document}

\title{
GAUGE-INVARIANT MESON PHOTOPRODUCTION\\WITH EXTENDED NUCLEONS
}

\author{H. HABERZETTL, C. BENNHOLD}
\address{Center for Nuclear Studies, Department of Physics,\\The George Washington University, Washington, D.C. 20052, U.S.A.}
\author{T. MART}
\address{Jurusan Fisika, FMIPA, Universitas Indonesia, Depok 16424, Indonesia}
\author{T. FEUSTER}
\address{Institut f\"ur Theoretische Physik, Universit\"at Gie{\ss}en, Gie{\ss}en, Germany}


\maketitle

\abstracts{The general gauge-invariant photoproduction formalism given by Haberzettl is 
applied to kaon photoproduction off the nucleon at the tree level, with form factors describing composite nucleons. 
Numerical results show that this gauge-invariance procedure, when compared to Ohta's, leads to a much improved 
description of experimental data. Predictions for the new Bonn SAPHIR data for $p(\gamma,K^+)\Lambda$ 
are given.}

\section{Formalism}

\begin{figure}[b!]%
\centerline{\psfig{file=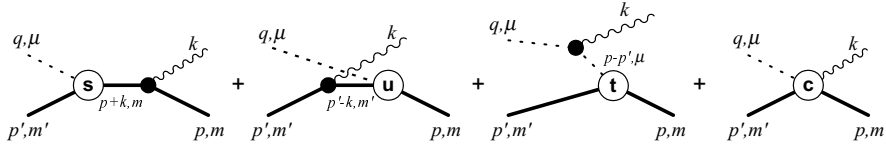,width=0.98\columnwidth,clip=,silent=}} 
\caption[fig1]{\label{tree} Tree-level photoproduction diagrams. Time proceeds
from right to left. 
The form factors $F_s$, $F_u$, and $F_t$ in the text describe
the vertices labeled $s$, $u$, and $t$, respectively, with appropriate
momenta and masses shown for their legs. The right-most diagram
corresponds to the contact-type interaction current required to restore gauge invariance.}
\end{figure}

Gauge invariance is one of the central issues
in dynamical descriptions of how photons interact with 
hadronic systems (see Refs.$\,$\cite{ohta89,hh97g}, and references therein).
For the simple example of 
$\gamma p \rightarrow n \pi^+$ with pseudoscalar coupling for the $\pi NN$ vertex,
one finds already at the tree level (see Fig.\ \ref{tree})%
that the corresponding
amplitude violates gauge invariance if the baryon structure is described by form factors.

However, it has been shown in Ref.$\,$\cite{hh97g} that gauge invariance can be restored
quite straightforwardly by adding a contact-type interaction current to the usual tree-level
$s$-, $u$-, and $t$-channel diagrams (cf.\ Fig.\ \ref{tree}). Indeed, one finds that the
amplitude then may be written  as a linear combination,$\,$\cite{hh97g,hh98k}
\begin{equation} \varepsilon \cdot \widehat{M}_{fi}=\sum_{j=1}^{4} \widehat{A}_j \overline{u}_n
\left(\varepsilon_\mu M_j^\mu \right) u_p \;\;, \label{mviol}
\end{equation}
of individually gauge-invariant currents,
\begin{eqnarray}
M_1^\mu &=& -\gamma_5 \gamma^\mu \; k \cdot \gamma \;\;, \label{m1}\\
M_2^\mu &=& 2\gamma_5 \left( p^\mu \; k\cdot p' - p'^\mu k\cdot p \; \right)  \;\;, \label{m2}\\
M_3^\mu &=&  \gamma_5 \left( \gamma^\mu \; k\cdot p  - p^\mu  \; k\cdot \gamma  \right)  \;\;, \label{m3}\\
M_4^\mu &=&  \gamma_5 \left( \gamma^\mu \; k\cdot p' - p'^\mu \; k\cdot \gamma  \right)  \;\;, \label{m4}
\end{eqnarray}
thus providing a manifestly gauge-invariant total current. The coefficient functions,
\begin{eqnarray}
\widehat{A}_1 &=&  \frac{ge}{s-m^2}\left( 1+\kappa_{\rm p}\right) F_s 
       + \frac{ge}{u-m'^2}\kappa_n F_u \;\;,\label{ah1}\\
\widehat{A}_2 &=&  \frac{2ge}{(s-m^2)(t-\mu^2)} \widehat{F} \;\;,\label{ah2}\\
\widehat{A}_3 &=&  \frac{ge}{s-m^2}\frac{\kappa_p}{m} F_s \;\;,\label{ah3}\\
\widehat{A}_4 &=&  \frac{ge}{u-m'^2}\frac{\kappa_n}{m'} F_u \;\;,\label{ah4}
\end{eqnarray}
depend on the respective form factors describing the three kinematical situations shown in Fig.\ \ref{tree}, i.e.,
\begin{eqnarray}
F_s &=& F_s(s) \;=\; f\!\left((p+k)^2,m'^2,\mu^2\right) \;\;, \label{f1}\\
F_u &=& F_u(u) \;=\; f\!\left(m^2,(p'-k)^2,\mu^2\right)\;\;, \label{f2}\\
F_t &=& F_t(t) \;=\; f\!\left(m^2,m'^2,(p-p')^2\right)  \;\;, \label{f3}
\end{eqnarray}
where $s$, $u$, and $t$ here are the Mandelstam variables. Putting
%
\begin{equation}\label{noff}
\mbox{Point-like nucleons: \quad}F_s= F_u=F_t=\widehat{F}=1  \;  
\end{equation}
corresponds to the case of structureless, point-like nucleons given by only the first three
diagrams of Fig.\ \ref{tree}.

As mentioned, for the realistic case of composite nucleons, maintaining gauge-invariance
requires the addition of a contact current. This is the origin of the function $\widehat{F}$ 
appearing here in $A_2$.  As it turns out,$\,$\cite{hh98k} there is considerable freedom in choosing $\widehat{F}$; 
in other words, we may use the particular form of $\widehat{F}$ to distinguish between 
different prescriptions for repairing gauge invariance.

One of the most popular prescriptions of this kind is due to Ohta$\,$\cite{ohta89}.
Using analytic continuation and minimal substitution, 
Ohta finds that the required $\widehat{F}$ is constant,
\begin{equation}  
  \mbox{Ohta: \quad} \widehat{F} =f\!\left(m^2,m'^2,\mu^2\right)= 1\;\;,   
\label{ohtaf}
\end{equation}
determined by the normalization condition for the form factor in the unphysical region where all 
three legs are on-shell. This corresponds precisely to what one obtains for $\widehat{F}$ in
the structureless case, Eq.\ (\ref{noff}),  and therefore the purely electric term $\widehat{A}_2$ 
of Eq.\ (\ref{ah2}) is treated as in the bare case, thus effectively freezing all degrees of freedom 
arising from the compositeness of the $\pi NN$ vertex.

The general meson photoproduction theory of Ref.$\,$\cite{hh97g} provides another, more flexible, way of choosing
$\widehat{F}$. Haberzettl's formalism allows one to take
$\widehat{F}$ as a linear combination
of all form factors appearing in the problem, i.e.,
\begin{equation}
\mbox{Haberzettl: \quad}
\widehat{F} = a_s F_s(s)+a_u F_u(u)+a_t F_t(t)\;\;,\label{hhb}
\end{equation}
where the coefficients are restricted by $a_s+a_u +a_t=1$ in order to provide the proper limit
for vanishing photon momentum (see Ref.$\,$\cite{hh98k} for details).

\section{Results}

We have tested the relative merits of both prescriptions 
for repairing gauge invariance 
for the kaon photoproduction reactions $\gamma p \rightarrow \Lambda K^+$ and $\gamma p \rightarrow \Sigma^0 K^+$.
In both cases, one can take over Eqs.\ (\ref{mviol})  and (\ref{ah1})-(\ref{ah4})  
by replacing the pion by $K^+$ and the
neutron by the respective hyperon. 

Using the resonance model of Ref.$\,$\cite{terry},
one of the main numerical results is summarized in Fig.\ 2. The upper panel shows 
$\chi^2$ per data point as a function of one of the leading Born coupling constants,
$g_{K \Lambda N} / \sqrt{4 \pi}$, for the two 
different gauge prescriptions by
Ohta and Haberzettl ($g_{K \Lambda N}$ was chosen here because $\chi^2$ shows very little
sensitivity on the other leading coupling constant, $g_{K \Sigma N}$ \cite{hh98k}).  
Clearly, Haberzettl's method provides $\chi^2$ values better than Ohta's by at least a factor of two, 
which, moreover, are almost independent of $g_{K \Lambda N}$, in stark contrast to Ohta's. 
In the fits the form factor cutoff $\Lambda$ was allowed
to vary freely. As is seen in the lower panel of Fig.\ 2, in the case of
Haberzettl's method, the cutoff decreases with
increasing $K \Lambda N$ coupling constant, leaving the magnitude of the
{\it effective} coupling, i.e., coupling
constant times form factor,  roughly constant. 
Since Ohta's method
does not involve form factors for electric contributions [cf.\ Eqs.\ (\ref{ah2}) and (\ref{ohtaf})] no such
compensation is possible there, and as a consequence
the cutoff remains insensitive to the coupling constant (see Ref.$\,$\cite{hh98k} for more details).

Figure 3 shows differential cross sections for $p(\gamma, K^+)\Lambda$ for four energies
for which new Bonn  SAPHIR data exist. The numerical results were
obtained within the coupled-channels
$K$-matrix model of Feuster and Mosel$\,$\cite{feuster98}.
The new data have {\it not} been included in the fit and the curves shown in Fig.\ 3 are, therefore, 
{\it predictions}. It is evident here that the method put forward by Haberzettl yields results
in better agreement with the experimental data. 

Our overall conclusion from the present findings is that Ohta's approach seems too restrictive
to account for the full hadronic structure while properly maintaining gauge invariance,
whereas the method put forward in Ref.$\,$\cite{hh97g} seems well capable of providing this facility.
This favorable conclusion regarding Haberzettl's method is corroborated also by the findings of 
Feuster and Mosel.$\,$\cite{feuster98}

This work was supported in part by Grant No.\ DE-FG02-95ER40907 of the U.S. Department of 
Energy.

\begin{figure}[t!]
\begin{minipage}[t]{.31\columnwidth}
\epsfig{file=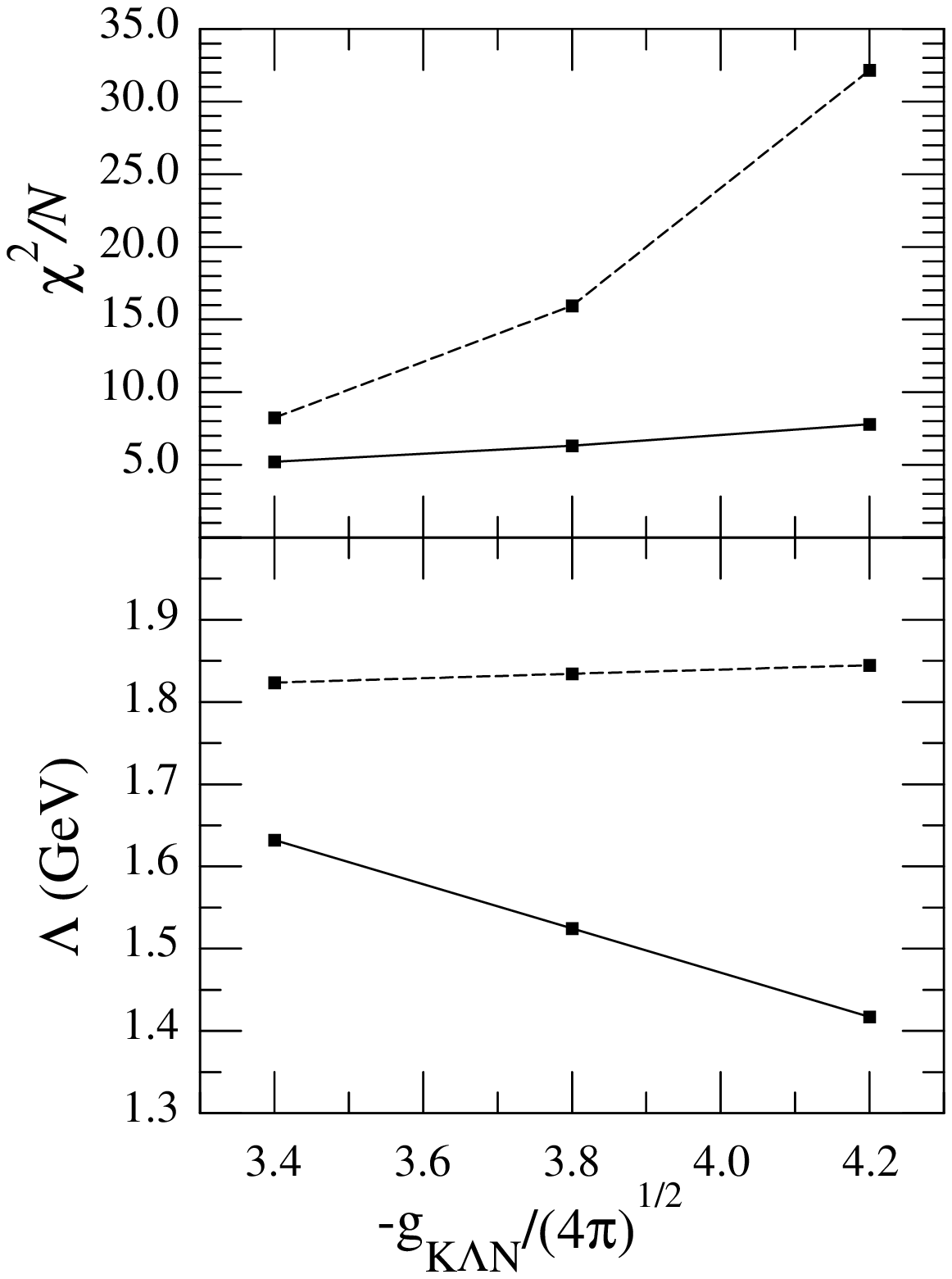,height=50mm,clip=,silent=}
\end{minipage}\hfill\begin{minipage}[t]{.66\columnwidth}
\epsfig{file=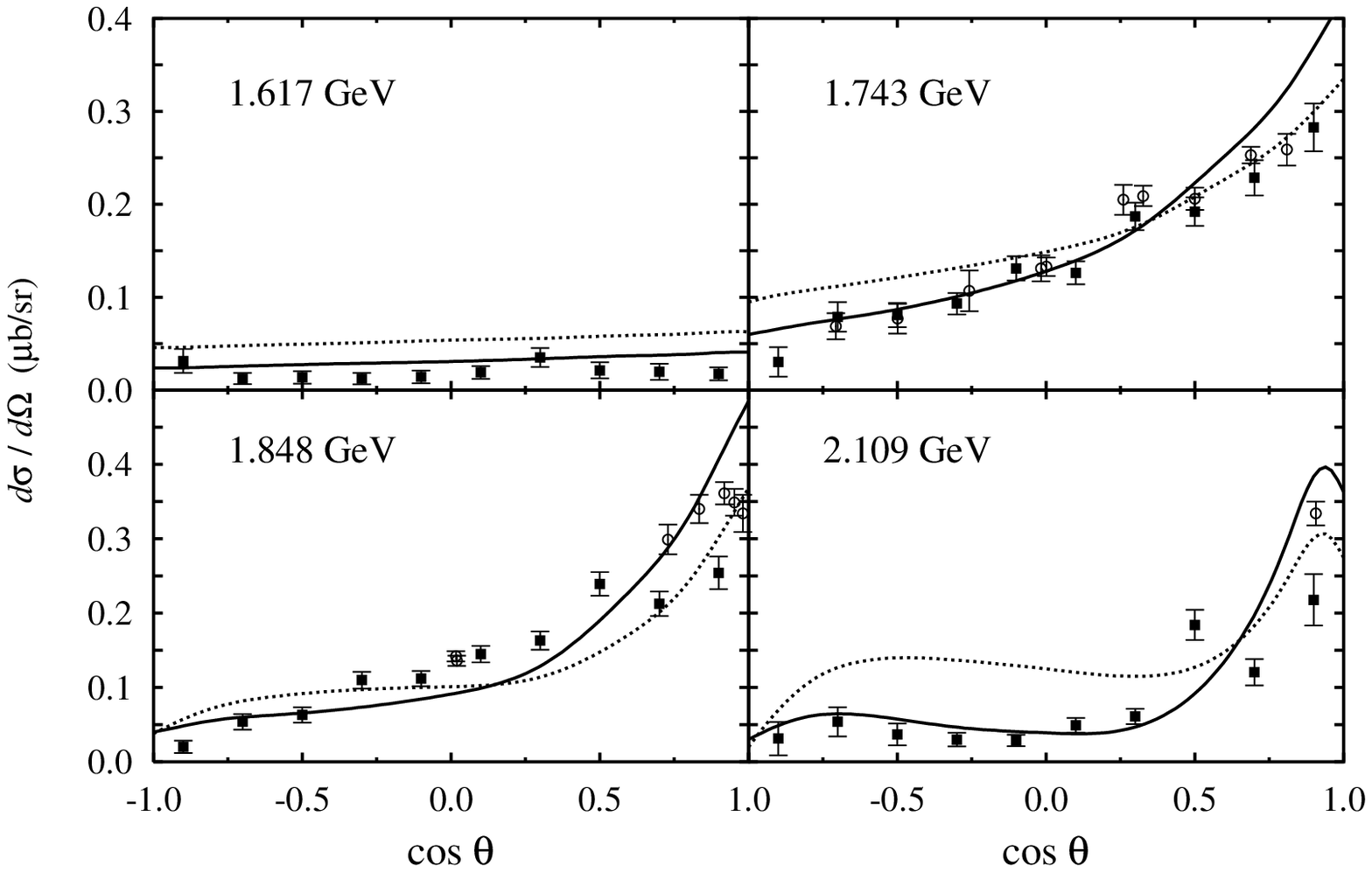,height=50mm,clip=,silent=}
\end{minipage}\\
\begin{minipage}[b]{0.36\columnwidth}
\caption{\label{chi}  $\chi^2/N$ and cutoff parameter $\Lambda$ as functions of
coupling constant  $g_{K\Lambda N}$ [solid
lines: Eq.\ (\ref{hhb}); dotted lines: Eq.\ (\ref{ohtaf})].}
\end{minipage}\hfill\begin{minipage}[b]{0.6\columnwidth}
\caption{\label{saphir} Differential cross sections for $p(\gamma, K^+)\Lambda$ [solid
lines: \protect{Eq.\ (\ref{hhb})}; dashed lines: Eq.\ (\ref{ohtaf}); experimental points:
old SAPHIR data$\,$\protect\cite{saphir94} (open circles); new SAPHIR data$\,$\protect\cite{saphir98} (solid squares)]. }
\end{minipage}
\end{figure}

\end{document}